\documentclass[10pt,twocolumn,preprintnumbers,superscriptaddress,nofootinbib,aps,prd]
{revtex4-2}

\usepackage{graphicx}
\usepackage{amsmath}
\usepackage{srcltx}
\usepackage[utf8]{inputenc}
\usepackage[colorlinks]{hyperref}
\usepackage{url}
\usepackage{amssymb}
\newcommand{\vev}[1]{\langle {#1} \rangle}
\newcommand{\lsim}{\lesssim}
\newcommand{\gsim}{\gtrsim}

\newcommand{\eq}[1]{Eq.~(\ref{#1})}

\newcommand{\ord}[1]{\mathcal{O}{(#1)}}
\newcommand{\beq}{\begin{equation}}
\newcommand{\eeq}{\end{equation}}
\newcommand{\bea}{\begin{eqnarray}}
\newcommand{\eea}{\end{eqnarray}}

\newcommand{\appropto}{\mathrel{\vcenter{
  \offinterlineskip\halign{\hfil$##$\cr
    \propto\cr\noalign{\kern2pt}\sim\cr\noalign{\kern-2pt}}}}}

\begin{document}

\pagestyle{plain}

\title{\boldmath Stellar Signals of a Baryon-Number-Violating Long-Range Force}

\author{Hooman Davoudiasl}
\email{hooman@bnl.gov}


\affiliation{High Energy Theory Group, Physics Department \\ Brookhaven National Laboratory,
Upton, NY 11973, USA}


\begin{abstract}

We entertain the novel possibility that long range forces may lead to violations of accidental symmetries, in particular baryon number.  Employing an ultralight scalar, with a mass $\ll$~eV, we illustrate that this scenario can lead to vastly disparate nucleon lifetimes, in different astronomical objects.   Such a long range interaction can yield a number of potentially observable effects, such as a flux of neutrinos at $\gsim 10$~MeV from the Sun and heating of old neutron stars.  We examine the prospects for constraining this scenario, with current and future astrophysical data, and find that neutron star heating provides the strongest present and near term bounds.  Simple extensions of our setup allow for the ultralight scalar to constitute the dark matter of the Universe.  This suggests that matter-enhanced baryon number violation can be a signal of ultralight dark matter, which has apparently been overlooked, so far.

\end{abstract}
\maketitle


\section{Introduction}

Experimental bounds on the lifetime of the proton imply that baryon number violation 
is extremely suppressed.  While it is not known if there are ultraviolet (UV) mechanisms that allow protons to decay, a variety of extensions of the Standard Model (SM) provide such a possibility, which leads to severe constraints on those models.  The requirements of gauge invariance within the SM do not allow renormalizeable interactions that violate baryon number, carried by quarks, making it associated with an {\it accidental} symmetry at low energies.  This circumstance then provides a potential explanation for the longevity of the proton, by relegating baryon number violation to non-renormalizeable operators suppressed by large mass scales, not far from the Planck mass $M_P \approx 1.2 \times 10^{19}$~GeV, where quantum gravity is expected to emerge.  

For example, let us consider the dimension-6 interaction 
\beq
O_6=\frac{(u u d\, \ell)_R}{M^2}\,,
\label{O6}
\eeq       
where $u$ and $d$ denote the up and down quarks, respectively, and $\ell=e,\mu$ is either and electron or a muon.  Here, $R$ denotes all right-handed fields and $M$ is a large effective scale  that encodes various UV physics couplings and masses.    
The above operator can lead to 
proton decay via $p\to \pi^0 \ell^+$, among other channels, where $\pi^0$ denotes the neutral pion.  The current bound on this partial lifetime is $\tau > 1.6\, (0.77)\times 10^{34}$~yr, for $e^+$ $(\mu^+)$, at 90\% confidence level \cite{ParticleDataGroup:2022pth}.  A rough order-of-magnitude estimate, assuming hadronic scales $\sim$~GeV, then yields $M\gsim 10^{16}$~GeV.  It is interesting that this is the right order of magnitude for the typical energy scale of grand unified theories \cite{Georgi:1974sy,Pati:1974yy,Dimopoulos:1981yj}, where baryon number violating processes mediated by states of mass $\sim M$ arise.

The standard picture of proton decay can be modified if we allow for new ``dark" states to couple to the SM quarks through non-renormalizeable interactions.  If such particles are light and can be emitted in the decay process, one can end up with new final states and kinematics that is affected by the mass of the dark particles \cite{Davoudiasl:2013pda,Davoudiasl:2014gfa,Fornal:2018eol,Helo:2018bgb,Barducci:2018rlx,McKeen:2020zni,Elahi:2020urr,Heeck:2020nbq,Fajfer:2020tqf}.  In this work, we take this possibility to an extreme limit: What if baryon number violating operators contain ultralight scalars?  Such fields may arise in the context of UV frameworks like string theory (see, for example, Ref.~\cite{Nusser:2004qu}).  Ultralight scalars can also provide suitable candidates for DM, and we will briefly discuss this possibility in the context of our scenario, near the end of this paper.  We note that in extreme regimes, ultralight bosons can potentially address certain features of astrophysical data that pose a challenge for conventional DM models that are based on weakly interacting massive particles \cite{Hu:2000ke,Hui:2016ltb}.

Let us consider the interaction in \eq{O6}, augmented by an ultralight scalar $\phi$, now as a dimension-7 operator
\beq
O_7 = \frac{\phi\,(u u d\, \ell)_R}{\Lambda^3}\,,
\label{O7}
\eeq
with $\Lambda$ some new large mass scale that depends on the details of the UV theory.  For our main discussion, we will only assume the above effective theory description, but we will come back to the question of UV completion later and briefly outline a possible setup for it.  

At low energies, where light quarks are confined, $O_7$ can lead to the emergence of an interaction of the type $\phi \,{\bar \ell^c} p$ that mediates baryon number violation.\footnote{Such a coupling was considered in Ref.~\cite{McKeen:2020zni}, but in a different regime and context.}  As we are interested in $\phi$ masses $m_\phi \ll$~eV, this low energy coupling can lead to proton decay through $p\to \ell^+ \phi$, which needs to be very suppressed in order to comply with extant bounds.  In addition, if $\phi$ is coherently {\it sourced} by ordinary matter, through coupling to nucleons or electrons, it can provide an effective mixing between $p$ and $\ell^+$, like the dimension-6 operator $O_6$, allowing the proton to decay through $p\to \pi^0 \ell^+$, for example.  However, depending on the background value of $\phi$ at the source, nucleon decay from $O_7$ may be enhanced in dense and large astronomical bodies.  This is a novel possibility that we will consider in this work.      
     
\section{Basic Model}  

We will assume the following low energy renormalizeable coupling for $\phi$ to nucleons 
\beq
g_N \phi \bar N\, N\,,
\label{phiNN}
\eeq
where $g_N$ is the coupling constant, $N=p, n$ is a nucleon, and $n$ denotes a neutron.  Current data yield the constraint $g_N \lsim 8.0\times 10^{-25}$, at $2\sigma$  \cite{Fayet:2017pdp,MICROSCOPE:2022doy}.  As mentioned before, $\phi$ is assumed to have a tiny non-zero mass.  We will consider $m_\phi \ll 10^{-14}$~eV, so that $\phi$ is effectively massless over distances of order the Earth radius $R_\oplus \approx 6400~\text{km} \approx (3\times 10^{-14}~\text{eV})^{-1}$.  The only other interaction beyond the SM we will consider is given by the above $O_7$ effective operator and we will limit our discussion to the case $\ell=e$, for simplicity.  Considering additional operators with other chiral and flavor structures will not change the main physics points we would like to emphasize here.  

We will use the formalism developed in Ref.~\cite{Claudson:1981gh} to derive the requisite hadron level  baryon-number-preserving and violating interactions, in our setup.  Baryon number preserving couplings of nucleons and mesons are given by 
\bea \nonumber
{\cal L}_{\rm P} &=& \left[\frac{(3F-D)}{2\sqrt{3}f_\pi}\,\partial_\mu \eta +
\frac{(D+F)}{2 f_\pi}\,\partial_\mu \pi^0\right] \bar p \gamma^\mu \gamma_5 p \\ 
&+&\frac{(D+F)}{\sqrt{2}f_\pi}\,\partial_\mu \pi^+ \, \bar p \gamma^\mu \gamma_5 n + \ldots\,, 
\label{LP}
\eea     
where we have written only terms relevant to the discussions below.  In the above Lagrangian, $D=0.80$, $F=0.47$ \cite{Aoki:2008ku}, and $f_\pi \approx 92$~MeV (note that our definition of the pion decay constant is smaller than that used in Ref.~\cite{Claudson:1981gh} by $\sqrt{2}$).  The baryon number violating hadronic interactions used in this work, corresponding to the dimension-7 operator in \eq{O7}, are 
\bea  \nonumber
{\cal L}_{\rm V} &=& \frac{\beta}{\Lambda^3}\, \phi\, \left[\overline{e^c_R}\, p_R -  
\frac{i}{2 f_\pi} (\sqrt{3} \eta + \pi^0)\overline{e^c_R} \,p_R \right] \\ 
&-& \frac{\beta}{\Lambda^3}\, \phi\,
\left[\frac{i}{\sqrt{2} f_\pi} \pi^+ \overline{e^c_R} \,n_R\right]
 + \text{\small H.C.}\,,
\label{LV}
\eea 
where $\beta=0.012 \pm 0.0026$~GeV$^3$ \cite{Aoki:2008ku}.
  
The above interactions can lead to nucleon decay in a few ways: (i) by proton decay in to $\phi$ and $e^+$ mediated by the first term in \eq{LV}, (ii) in 3-body decays into $\phi$, a meson, and $e^+$ from \eq{LV}, (iii) via baryon number preserving emission of a meson by nucleons in \eq{LP} and proton-$e^+$ mixing from the first term in \eq{LV}, or (iv) through the point interactions (\ref{LV}) involving a nucleon, a meson, and $e^+$.  Here, we are assuming $\vev{\phi}\neq 0$ for (iii) and (iv).    

\section{Nucleon Decay Rates}

In what follows we can, to a good approximation, ignore the mass of the positron: $m_e\to 0$.  One can then show that, in our setup, the left-handed proton and positron mix through the off-diagonal mass term 
\beq
\mu = \kappa \vev{\phi}
\label{mu}
\eeq
with $\kappa \equiv \beta/\Lambda^3$, leading to a mixing angle (in the mass basis) given by  
\beq
\sin\xi \approx -\frac{\mu}{m_p}\,,
\label{xi}
\eeq
where $m_p$ is the proton mass.  The decay width for $p \to \phi \, e^+$ is given by 
\beq
\Gamma(p \to \phi \, e^+) = \frac{\kappa^2}{32 \pi}\,m_p\,.
\label{Gamphi}
\eeq
Note that 3-body nucleon decays are generally more suppressed compared to 2-body decays, by factors $\gsim \ord{10}$ \cite{Claudson:1981gh}.  Hence, we will base our phenomenological bounds on constraints from the dominant 2-body decays, in what follows.  This suffices to illustrate key phenomenology and typical sizes of the new  effects, at the level of numerical accuracy intended for our work.

For the decay modes $p \to {\cal M} e^+$, where the meson ${\cal M} = \pi^0, \eta$.   With our assumptions, we find
\beq
\Gamma(p \to {\cal M} e^+) = \frac{\lambda_{\cal M}^2}{32 \pi}\,m_p \left(1- \frac{m_{\cal M}^2}{m_p^2}\right)^2\,,
\label{GamM}
\eeq
where 
\beq
\lambda_\pi \equiv \frac{(D+F+1)\mu}{2 f_\pi}
\label{lampi}
\eeq
and 
\beq
\lambda_\eta \equiv \frac{(3F - D + 3)\mu}{2 \sqrt{3} \,f_\pi}\,.
\label{lameta}
\eeq
We also obtain the neutron decay width 
\beq
\Gamma(n\to \pi^- e^+) = \frac{\lambda_\pi^2}{16 \pi}\,m_n \left(1- \frac{m_{\pi^-}^2}{m_n^2}\right)^2.
\label{neutron-decay}
\eeq

\section{Nucleon Decay in ``Empty Space"}

Since $|\lambda_{\cal M}|\sim \ord{\mu/f_\pi}$, we find that 
\beq
\frac{\Gamma(p \to \phi \, e^+)}{\Gamma(p \to {\cal M} e^+)}
\sim \left(\frac{f_\pi}{\vev{\phi}}\right)^2\,.
\label{ratio}
\eeq
The above implies that in empty space, or if $g_N\to 0$, the 2-body decay channels originating from the interaction in \eq{O7} that include a final state $\phi$ would dominate.  However, for $\vev{\phi}\gg f_\pi$, these modes could be sub-dominant.  To see this, let us fix $g_N = 10^{-25}$; this will be our reference value, unless otherwise specified. 

For $m_\phi^{-1}$ large compared to size $R_*$ of astronomical bodies with mass $M_*$, we roughly have 
\beq
\phi_* \approx -\frac{g_N (M_*/m_N)}{4 \pi \,R_*}\,, 
\label{phi*}
\eeq
near the surface of the body (and a similar magnitude within it); the sign of the background $\phi$ corresponds to an attractive force between nucleons.  We will set $m_\phi=10^{-16}$~eV, which is of order the inverse radius of the Sun, $R_\odot \approx 7.0 \times 10^5$~km.  Then, for the Earth, with $M_\oplus \approx 6.0\times 10^{27}$~g, we get $\vev{\phi}_\oplus \sim 10^4 f_\pi$.  This suggests that $p \to \phi \, e^+$ would have negligible effect on the nucleon lifetime, near astronomical bodies.  However, for our reference value of $g_N$, this is not the case away from stars and planets, {\it i.e.} ``empty space."  To see this, note that the local Galactic density of nucleons, about 1 per cm$^3$, would yield $\vev{\phi}_{\rm G}\sim 10^{-7}$~eV$\ll f_\pi$; the cosmic nucleon density is far below the local value resulting in a much smaller induced $\vev{\phi}$.

The preceding analysis suggests that in empty space, or if $g_N = 0$, the decay mode $p \to \phi \, e^+$ dominates the nucleon lifetime in our scenario.  In this case, if no other effective operators besides $O_7$ are important, neutron decay will go through a 3-body decay, which would be comparatively more suppressed.  If these decays are too fast, they can lead to an excess diffuse background photon flux and conflict with astrophysical data.  Constraints on the lifetime of GeV scale dark matter (DM), rescaled by a factor of $\sim 5$ to reflect the smaller number density of cosmic baryons, would give $\tau_N \gsim 10^{23}$~s \cite{Bell:2010fk}.  Yet, this bound is far less stringent than those obtained from nucleon decay experiments, as explained below.

\section{Laboratory Constraints}

If nucleon decay is dominated by $O_7$ in \eq{O7}, then in the limit $g_N\to 0$, which also applies to empty space, $p \to \phi \, e^+$ is constrained by $\tau(p\to e^+ X) > 7.9 \times 10^{32}$~yr, at 90\% confidence level (CL), where $X$ is assumed to be massless \cite{ParticleDataGroup:2022pth}. Using Eq.~(\ref{Gamphi}), we then find
\beq
\Lambda \gsim 6\times 10^9~\text{GeV} \quad ;\quad \text{Earth, $g_N=0$}.
\label{Lam-gN0}     
\eeq

The bound (\ref{Lam-gN0}) applies as a minimum requirement in our scenario.  However, for our reference value of $g_N$ more stringent constraints can be derived, due to the value of 
\beq
\vev{\phi}_\oplus \approx 8.7 \times 10^{2}
\left(\frac{g_N}{10^{-25}}\right)~\text{GeV}\,,
\label{phiE}
\eeq 
induced by terrestrial  nucleons.  In this case, $p\to e^+ \pi^0$ yields the most stringent experimental limit on our model, with $\tau(p\to e^+ \pi^0) > 1.6 \times 10^{34}$~yr, at 90\% CL \cite{ParticleDataGroup:2022pth}.  Using \eq{GamM}, we find 
\beq
\Lambda \gsim 2\times 10^{11} \left(\frac{g_N}{10^{-25}}\right)^{1/3}~\text{GeV} \quad ;\quad \text{Earth, $g_N\neq 0$}.
\label{LamEarth}     
\eeq

Given the above bound from terrestrial experiments, we may inquire if more stringent constraints can result from observations of denser objects, inducing larger values of $\vev{\phi}$.    
We will consider two possibilities: First, nucleon decay in the Sun, leading to a flux of neutrinos at the Earth.  Secondly, we will consider how enhanced nucleon decay can lead to anomalous heating of old neutron stars. 

\section{Solar Neutrinos from Nucleon Decay} 

Fast nucleon decays in the Sun can lead to an anomalous flux of neutrinos, with energies exceeding $\sim 10$~MeV, at the Earth.  In fact, the Super-Kamiokande (SK) experiment has looked for such a flux and placed bounds on possible baryon number violation in the Sun \cite{Super-Kamiokande:2012tld}, mediated by monopoles \cite{Rubakov:1981rg, Rubakov:1982fp,Callan:1982ah,Callan:1982au} predicted in grand unified theories \cite{tHooft:1974kcl,Polyakov:1974ek}.  The SK analysis, based on 176 kton-yr of exposure \cite{Super-Kamiokande:2012tld}, focused on the emission of a $\pi^+$ in proton decay which results in the final states $\pi^+ \to \mu^+ \nu_\mu \to e^+ {\bar \nu_\mu} \nu_\mu \nu_e$.  

In the minimal model presented here, we do not have a prompt 2-body nucleon decay into $\pi^+$\footnote{As before, we are ignoring 3-body decays, such as $p\to \pi^+ \pi^- e^+$, as subdominant.  However, their numerical significance may not be completely negligible, once we take account of the branching ratio for $\eta \to \pi^+$ + `anything' following $p \to \eta\, e^+$ that we have considered in our analysis.  Nonetheless, we do not expect the contribution from  3-body decays to change our estimated bounds significantly.}.  However, the decay chain $p\to e^+ \eta$ can yield a $\pi^+$ via $\eta \to \pi^+\pi^-\pi^0$ and $\eta \to \pi^+\pi^- \gamma$, which together have a branching fraction ${\rm Br}(\eta\to \pi^+)\approx 27\%$ \cite{ParticleDataGroup:2022pth}.  In order to use the SK analysis more directly and also as a conservative approach, we will thus focus on $p\to e^+ \eta$ to constrain our model.  We will take the number of protons in the Sun ${\cal N}_p\approx 10^{57}$ \cite{Davoudiasl:2011sz} and adopt the BP2004 Solar model in Ref.~\cite{Bahcall:2004fg} for the mass density $\rho(r)$ of the Sun, where $r$ is the Solar radial coordinate.\footnote{Detailed numerical data for the BP2004 Solar model can be found at:  \url{http://www.sns.ias.edu/~jnb/SNdata/Export/BP2004/bp2004stdmodel.dat}.}

\begin{figure}[t]\vskip0.25cm
\centering
\includegraphics[width=\columnwidth]{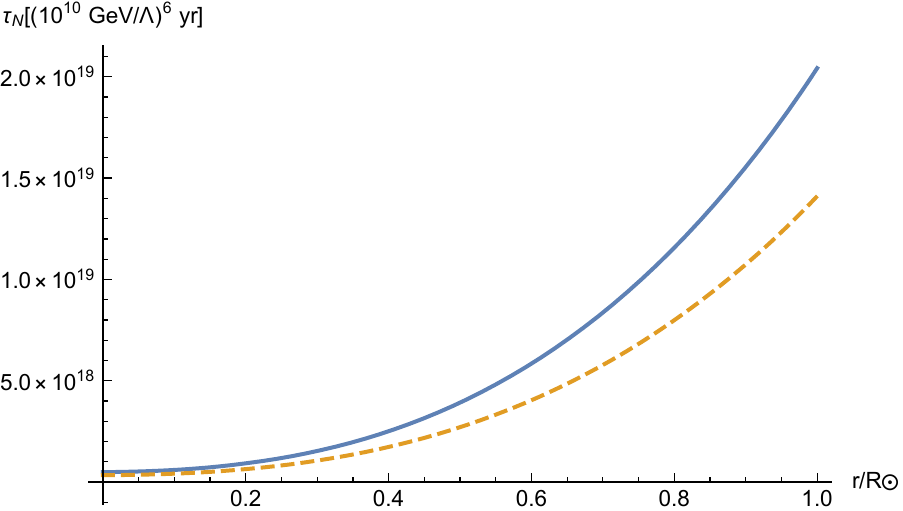}
\caption{Nucleon lifetimes in the Sun, as a function of solar radius.  The solid curve corresponds to the proton and the dashed curve to the neutron.}
\label{LHC-exp}
\end{figure}

We can calculate $\phi$ as a function of the radial distance $r_0=|\vec{r}_0|$ from the center of the Sun according to
\beq
\phi(r_0) = -\frac{g_N}{2 m_N}\int_0^{R_\odot}\!\!\!  dr\, r^2\, \rho(r)
\int_{-1}^{+1} \!\!\! dx\, \frac{e^{-m_\phi |\vec{r}-\vec{r}_0|}}{|\vec{r}-\vec{r}_0|}\,.
\label{phir0}
\eeq
where, by spherical symmetry, we have chosen $\vec{r}_0$ to lie on the $z$-axis and identified the angle between $\vec{r}$ and $\vec{r}_0$ as the polar angle $\theta$; here $x\equiv \cos\theta$.  The rate ${\cal R}_{\eta e}$ of proton decay $p\to \eta \,e^+$ over the whole Sun can then be calculated by the replacement $\vev{\phi}\to \phi(r)$ in \eq{GamM} and performing the following integral 
\beq
{\cal R}_{\eta e} = \frac{4 \pi}{m_N} \int_0^{R_\odot} \!\!\! dr\, r^2 
\rho(r)\,\Gamma(r)_{(p\to \eta \,e^+)}\,.
\label{Retae}
\eeq

The SK collaboration has placed a 90\% CL limit on the Solar neutrino flux,  $I_{90}=166.6$~cm$^{-2}$ s$^{-1}$ \cite{Super-Kamiokande:2012tld}, coming from the monopole catalyzed $p\to \pi^+$ + `anything' decay.  Adapting the SK analysis using the rate in \eq{Retae} above, we obtain 
\beq
{\cal R}_{\eta e} = 
\frac{4 \pi\, d_{\rm AU}^2 I_{90}}{3 {\rm Br}(\eta\to \pi^+)\,(1 - a_{\pi^+})}\,,
\label{rate90CL}
\eeq
where $d_{\rm AU} \approx 1.5 \times 10^8$~km is the radius of the Earth orbit around the Sun and 
$a_{\pi^+} = 0.2$ is the absorption probability for $\pi^+$ in the Solar center, which we adopt for the whole volume, as an approximation.  The above then yields the bound 
\beq
\Lambda \gsim 2 \times 10^{10}\left(\frac{g_N}{10^{-25}}\right)^{1/3}~\text{GeV} 
\quad ; \quad \text{Solar $\nu$ flux}\,, 
\label{Sol-bound}
\eeq
on $\phi$-enhanced proton decay in the Sun.  This is a conservative result and we expect that including additional channels can somewhat enhance this bound. However, we still expect that the bound obtained in \eq{LamEarth} is stronger even with a refinement of the SK analysis, given the $\Lambda^{-6}$ dependence of the neutrino flux.  Similarly, improvements of \eq{Sol-bound} with roughly twice the SK exposure available \cite{Nakano:2020fgc} -- compared to that used in Ref.~\cite{Super-Kamiokande:2012tld} -- are expected to be modest \cite{Hu:2022wcd}.  

\section{Neutron Star Heating}

Let us now examine how $\phi$-enhanced decays of nucleons in a neutron star (NS) can be constrained from bounds on anomalous heating effects.\footnote{For some recent work utilizing NS heating to probe new physics see, for example, Refs.~\cite{Baryakhtar:2017dbj,Acevedo:2019agu,McKeen:2021jbh}.}  Here, we give an approximate treatment and only consider neutrons in the star.  In our scenario, we then focus on $n\to \pi^- e^+$.  

We will take the typical NS mass $M_{\rm NS} \approx 1.5 M_\odot$, corresponding to number of neutrons $N_n\approx 2\times 10^{57}$, and radius $R_{\rm NS}\approx 10$~km.  We assume that the entire rest mass energy of the decaying neutron is deposited in the NS.  This is a fair approximation as the NS nucleon number density $\sim 4 \times 10^{38}$~cm$^{-3}$ and the cross section for $\sim 10$~MeV neutrino nucleon interactions $\sigma_{\nu N}\sim 10^{-42}$~cm$^2$ \cite{Formaggio:2012cpf} yields a neutrino mean free path of $\ord{10~\text{m}} \ll R_{\rm NS}$.  Hence, the weakest interacting final states resulting from $n\to \pi^- e^+$ decays will scatter many times before leaving the NS and can thus deposit a significant fraction of their energy.

We adopt a constant NS density 
\beq
\rho_{\rm NS} = \frac{M_{\rm NS}}{(4 \pi/3)R_{\rm NS}^3}\approx 7\times 10^{14}~\text{g} \text{cm}^{-3}\,, 
\label{rhoNS}
\eeq
following the parameter choices above.  For $r<R_{\rm SN}$, the scalar background within the star will then be given by
\beq
\phi_{\rm NS}(r) \approx -\frac{g_N \,\rho_{\rm NS}}{6\, m_n}\, R_{\rm NS}^2
\left(3 - \frac{r^2}{R_{\rm NS}^2}\right).
\label{phiNS}
\eeq
The neutron decay rate $\Gamma_n^{\rm NS}$ over the entire NS is then given by
\beq
\Gamma_n^{\rm NS} = 4 \pi\, \frac{\rho_{\rm NS}}{m_n}
\int_0^{R_{\rm NS}} dr \, r^2 \, \Gamma(r)_{(n\to \pi^-\, e^+)}\,,
\label{GamNS} 
\eeq
where the $\phi$ dependence in \eq{neutron-decay} is replaced with $\phi_{\rm NS}(r)$.  Here, we note that the size of the $\phi$ potential induced within the NS would be $\ord{10^9}$ times larger than that of the Earth in \eq{phiE}.  Given our reference parameters, we then expect that the relative shift of the nucleon mass caused by the attractive potential would be $\delta m_N/m_N \sim 10^{-13}$, which is completely negligible in our discussion.         

In the steady state, the heat deposition rate and the black body radiation balance each other and we thus get 
\beq
m_n \Gamma_n^{\rm NS} \approx 4 \pi R_{\rm NS}^2 \,\sigma_{\rm SB} T_{\rm NS}^4\,,
\label{SteadyState}
\eeq           
where $\sigma_{\rm SB} = \pi^2/60$ is the Stefan-Boltzmann constant and $T_{\rm NS}$ is the surface temperature of the star.  The coldest known neutron star is the pulsar PSR J2144–3933 -- estimated to be $\sim 3 \times 10^8$~yr old -- which has a temperature $T_{NS} < 42000$~K, based on data from the {\it Hubble Space Telescope} (HST) \cite{Guillot:2019ugf}.  It is estimated that an NS of this age would cool to $T_{\rm NS}\sim \ord{100~\text{K}}$ \cite{Yakovlev:2004iq} without additional  heating mechanisms.  Using the above bound on $T_{\rm NS}$ together with \eq{SteadyState}, we obtain 
\beq
\Lambda \gsim 7 \times 10^{11}\left(\frac{g_N}{10^{-25}}\right)^{1/3}~\text{GeV} \quad; \quad \text{NS (HST).}
\label{HST}
\eeq          

The lower bound on $\Lambda$ in \eq{HST} is stronger than the ones from terrestrial proton decay and Solar neutrino data.  However, it is expected that the bound on $T_{\rm NS}$  could be strengthened from upcoming measurements \cite{Baryakhtar:2022hbu}.  For example, Ref.~\cite{Chatterjee:2022dhp} finds that for an NS within 10~pc of the Earth, $T_{\rm NS}\gsim 2400$~K can be probed by the {\it James Webb Space Telescope} (JWST) \cite{Gardner:2006ky}, with modest observing times.  Such projected sensitivity would allow detection of anomalous NS heating for 
\beq
\Lambda \lsim 4\times 10^{12}~\left(\frac{g_N}{10^{-25}}\right)^{1/3}~\text{GeV} \quad ;\quad \text{NS (JWST).}
\label{JWST}
\eeq

We end this discussion by asking whether the number of neutrinos released by the neutron decay could be larger than $\ord{10^{58}}$, corresponding to the $\ord{M_\odot}$ energy budget of a core collapse supernova emitted in $\sim 10-100$~MeV neutrinos.  Using the bound in \eq{HST}, we estimate that a neutron star would have a rate of emission of neutrinos $\sim 10^{30}$~s$^{-1}$.  Over a time of a few billion years, we then expect $\ord{10^{47}}$ neutrinos to result form the enhanced decay rate of neutrons, which is a negligible $\ord{10^{-11}}$ fraction of that released in a core collapse supernova.  Hence, the neutron decay in the NS would not yield an anomalous diffuse neutrino background that is detectable above the standard expectation.

\section{Constraints from Early Universe}

The high density of baryons in the early Universe can source a large value for $\phi$.  One may then ask whether this could lead to very fast nucleon decay, possibly causing conflict with cosmological data.  We will address this question next.  

In the radiation dominated era the size of the horizon scales with temperature as $T^{-2}$ and the density of baryons grows as $T^3$.  Hence, going to higher temperatures would suppress the value of $\phi$ as $T^{-1}$.  Since $\phi \approx - g_N n_N m_\phi^{-2}$, for the maximal effect we consider the temperature corresponding to a horizon size $\sim m_\phi^{-1}$.  This roughly corresponds to $T\sim$~MeV, which is also characteristic of the Big Bang Nucleosynthesis (BBN).  

Given the baryon asymmetry of $\sim 10^{-10}$, the nucleon density is given by        
$n_N\sim 10^{-10} g_* T^3\sim  10^{-9} T^3$, where $g_*\sim 10$ counts the relativistic degrees of freedom.  We thus find $|\vev{\phi}_{\rm BBN}| \sim 10^7$~GeV.  To make sure that the BBN is not disrupted, we require that the neutron lifetime obtained from \eq{neutron-decay} is much longer than the experimental value of the free neutron lifetime $\approx 878$~s \cite{ParticleDataGroup:2022pth}, which enters calculations of elemental abundances.  We find that $\Lambda \gsim 2\times 10^6$~GeV would satisfy our requirement for the BBN.  However, the above bound is far less stringent than those we obtained from other constraints.

\section{Possible UV Framework}

Let us now briefly discuss potential UV theories that can lead to the 
effective interaction in \eq{O7}.  Our preceding discussions implicitly assumed that other possible interactions are not the dominant source of nucleon decay.  In particular, our phenomenology is predicated on the assumption that $M \gg \Lambda$, otherwise the effect of the dimension-7 operator $O_7$ would generally be negligible compared to that from the dimension-6 interaction $O_6$.  We will roughly sketch a possible setup that would realize this expectation, below.

A fairly simple possibility for justifying our assumptions is to postulate a  $\mathbb{Z}_2$ parity that acts on $\phi$ and leptons: $\mathbb{Z}_2(\phi) = \mathbb{Z}_2(e) = -1$, with all other SM fields having positive parity.  Such a charge assignment would then forbid $O_6$, while allowing $O_7$\footnote{We note that quantum gravity effects can potentially violate the assumed $\mathbb{Z}_2$ parity.  In that case, the scale $M$ of $O_6$ in \eq{O6} may be pushed to the Planck mass $M_P$, making its contribution to nucleon decay negligible in our analysis.}.  To generate the required operator we may, for example, introduce a parity-even vector-like field $\cal E$ that has the SM quantum numbers of $e_R$.  Then, we can have a coupling $y\, \phi\, \bar {\cal E}_L e_R$ and a dimension-6 operator
\beq
\frac{(uud\, {\cal E})_R}{\Lambda'^2}\,,
\label{new-dim6}
\eeq
consistent with the assumed $\mathbb{Z}_2$ symmetry.  One can write down a mass term for ${\cal E}$; the mass $m_{{\cal E}}$ needs to be above the weak scale to accommodate general experimental bounds on new particles that have electric charge.  One can then integrate ${\cal E}$ out and arrive at $O_7$ through \eq{new-dim6}.  Here, we have implicitly assumed that  $m_{{\cal E}}$ and the Yukawa coupling $y$ are chosen such that 
$\Lambda^3 =\Lambda'^2 m_{{\cal E}}/y$. 

A potential issue with the above setup is that the coupling of $\phi$ to nucleons in \eq{phiNN} is not consistent with the assumed $\mathbb{Z}_2$ symmetry.  One way to address this problem is to introduce a new parity-odd singlet scalar $\Phi$ with a vacuum expectation value $\vev{\Phi}\neq 0$ that spontaneously breaks the $\mathbb{Z}_2$ symmetry.  We may assume that there are heavy new vector-like fields ${\cal Q}$, with the gauge quantum numbers of a right-handed SM quark $q_R$, but odd under the $\mathbb{Z}_2$ parity.  Then, we may write down the couplings $\Phi \bar{\cal Q}_L q_R$ and $\phi\bar{\cal Q}_L q_R$.  Integrating out ${\cal Q}$ would lead to an effective operator $\propto \Phi \phi \bar q q$, where $q_R$ coupling with its left-handed partner is achieved through the SM mass term (and also non-perturbative QCD, in the case of  light quarks).  For $\vev{\Phi} \neq 0$, one can thus obtain the dimension-4 interaction $\phi \bar q q$, and hence a $\phi$ coupling to nucleons, as in \eq{phiNN}.  

We will not further discuss the details of the aforementioned setup.   In particular, the simple framework described above may entail tuned parameters, whose origin we will not elaborate upon, but could perhaps be explained with more model-building. We note that, in principle, one can choose the masses of the new fields ${\cal E}$ and ${\cal Q}$, which are charged under SM gauge interactions, to be within the reach of current or future collider experiments.  If so, one can achieve the required highly suppressed $\phi$-nucleon coupling $g_N\lsim 10^{-25}$ from presumably tiny Yukawa couplings of  $q$ and ${\cal Q}$ with the scalars $\phi$ and $\Phi$.                 

\section{Extension: Ultralight scalar DM}

One may easily extend the setup considered so far in order to make $\phi$ a viable DM candidate.  This requires that the initial misalignment of $\phi$ is chosen such that once it starts oscillating when the Hubble rate is close to $m_\phi$, it can constitute DM at $T\sim$~eV, near the time of matter-radiation equality.  The preceding analysis implies that for the chosen $\phi$ mass of $10^{-16}$~eV, it starts its coherent oscillations at $T\sim 1$~MeV.  Since the energy density $m_\phi^2 \phi^2$  of a coherently oscillating scalar redshifts as $T^3$, that is like matter, we see that an initial value $\phi_i \sim 10^{25}$~eV can furnish the correct DM contribution.

It is interesting to note that a thermal mechanism can set the desired $\phi_i$ quite naturally (see, for example, Refs.~\cite{Batell:2021ofv,Croon:2022gwq}).  At $T\sim$~MeV, $e^+e^-$ pairs are still in equilibrium in the early Universe and far more abundant than nucleons, with a number density $n_e\sim T^3$.  If we postulate a coupling for $\phi$ to electrons $g_e\approx 10^{-25}$, at the same level as for nucleons, a simple estimate yields $|\phi_i| \sim g_e n_e m_\phi^{-2}\sim 10^{25}$~eV, which is the right size for misalignment of $\phi$, according to the preceding estimates.  
The $2\sigma$ bound on long range forces coupled to electrons requires $g_e \lsim 1.4 \times 10^{-25}$ \cite{Fayet:2017pdp,MICROSCOPE:2022doy} and our choice is therefore consistent with current constraints.  

We point out that with the above value of $\phi_i$, induced by the thermal $e^+e^-$ population, the bound on neutron lifetime during BBN that we considered before gets enhanced to $\Lambda\gsim 2\times 10^9$~GeV.  This is still weaker than the main constraints we derived in our analysis.  Also, assuming a DM energy density of $\sim 0.3$~GeV cm$^{-3}$ \cite{ParticleDataGroup:2022pth} around the Solar System, we find an amplitude $\phi \sim 10^{13}$~eV for the ambient scalar field, about an order of magnitude larger than that generated by the Earth in \eq{phiE}.  This can lead to a slightly stronger bound on $\Lambda$, compared to that of \eq{LamEarth}, but not stronger than the NS heating bound.  Hence, making $\phi$ a viable DM candidate can be consistent with other potential signals we considered before.  

Whether one assumes both couplings $g_N$ and $g_e$ at the aforementioned levels, or only one of them, affects the phenomenology of the model in detail.  However, much of the discussion in our work will stay qualitatively the same, especially in environments like the Sun or the Earth where electrons and nucleons have similar number densities.  A more detailed discussion of the DM scenario and other choices of parameters is outside the scope of this work.

\section{Summary and Conclusions}        

In this paper, we considered the possibility that baryon number violating operators that result in nucleon decay may include an ultralight scalar field $\phi$.  Such a scalar can mediate a long range force acting on ordinary matter, {\it i.e.} nucleons or electrons, which would then source a background value for it.  In that case, one may have enhanced nucleon decay rates near and within astronomical bodies.  Using chiral perturbation theory, we calculated the rates for nucleon decay in this scenario, using a particular dimension-7 operator, as an example.  Both final states that include $\phi$, representing vacuum decays or zero matter coupling, or those that involve only SM decay products, with rates enhanced by matter density, were considered.  

We examined how this scenario can be constrained in a number of ways, including from cosmology, terrestrial laboratory data, as well as anomalous Solar neutrino flux and old neutron star heating, originating from faster nucleon decay in the stellar environment.  We found that the strongest constraint is currently obtained from the HST observations that provide an upper bound on the temperature of the coldest known neutron star.  Future data from the JWST can be significantly more constraining and lead to stronger bounds.  We also provided a sketch of how our setup can be realized in a UV model, which involves particles that can potentially be observable at high energy colliders, depending on their masses.  Finally, we briefly examined the possibility that the postulated ultralight scalar may constitute all of dark matter in the Universe, which can provide significant additional motivation for considering our scenario.

\vskip0.5cm
\begin{acknowledgments}
We thank P. Boyle and R. Szafron for helpful discussions.  This work is supported by the US Department of Energy under Grant Contract DE-SC0012704.
\end{acknowledgments}

\bibliography{bv-scalar}

\end{document}